\documentstyle[12pt,epsf,epsfig]{article}
\newfont{\thiplo}{msbm10 scaled\magstep 2}
\newfont{\gothic}{eufb10 scaled\magstep 2}
\newfont{\unc}{eurb10} 
\newskip\humongous \humongous=0pt plus 1000pt minus 1000pt
\def\caja{\mathsurround=0pt}\def\eqalign#1{\,\vcenter{\openup1\jot \caja
        \ialign{\strut \hfil$\displaystyle{##}$&$ 
        \displaystyle{{}##}$\hfil\crcr#1\crcr}}\,}
\newif\ifdtup

\def\eqright #1\cr{\noalign{\hfill$\displaystyle{{}#1}$}}
\def\eqleft #1\cr{\noalign{\noindent$\displaystyle{{}#1}$\hfill}}
\def\oldreffmt#1{\rlap{[#1]} \hbox to 2\parindent{}}

\def\figfmt#1{\rlap{Figure {#1}} \hbox to 1in{}}

\def\sectioneq{\def\theequation{\thesection.\arabic{equation}}{\let
\holdsection=\section\def\section{\setcounter{equation}{0}\holdsection}}}%

\newcounter{holdequation}

\def\begineq #1\endeq{$$ \refstepcounter{equation}\eqalign{#1}\eqno
	(\theequation) $$}
\def\contlimit{\,{\hbox{$\longrightarrow$}\kern-1.8em\lower1ex
\hbox{${\scriptstyle (a\rightarrow0)}$}}\,}
\def\centeron#1#2{{\setbox0=\hbox{#1}\setbox1=\hbox{#2}\ifdim
\wd1>\wd0\kern.5\wd1\kern-.5\wd0\fi
\copy0\kern-.5\wd0\kern-.5\wd1\copy1\ifdim\wd0>\wd1
\kern.5\wd0\kern-.5\wd1\fi}}
\def\centerover#1#2{\centeron{#1}{\setbox0=\hbox{#1}\setbox
1=\hbox{#2}\raise\ht0\hbox{\raise\dp1\hbox{\copy1}}}}
\def\centerunder#1#2{\centeron{#1}{\setbox0=\hbox{#1}\setbox
1=\hbox{#2}\lower\dp0\hbox{\lower\ht1\hbox{\copy1}}}}
\def\lsim{\;\centeron{\raise.35ex\hbox{$<$}}{\lower.65ex\hbox
{$\sim$}}\;}
\def\gsim{\;\centeron{\raise.35ex\hbox{$>$}}{\lower.65ex\hbox
{$\sim$}}\;}
\def\super#1{\ifmmode \hbox{\textsuper{#1}}\else\textsuper{#1}\fi}
\def\textsuper#1{\newcount\holdspacefactor\holdspacefactor=\spacefactor
$^{#1}$\spacefactor=\holdspacefactor}
\def\getcite#1,{\advance\citenumber by1
\def\getcitearg{#1}\def\lastarg{@}
\ifnum\citenumber=1
\ref{#1}\let\next=\getcite\else\ifx\getcitearg\lastarg\let\next=\relax
\else ,\ref{#1}\let\next=\getcite\fi\fi\next}
\def\pom{{\rm P\kern -0.53em\llap I\,}}
\def\spom{{\rm P\kern -0.36em\llap \small I\,}}
\def\sspom{{\rm P\kern -0.33em\llap \footnotesize I\,}}

\relax
\def\contlimit{\,{\hbox{$\longrightarrow$}\kern-1.8em\lower1ex
\hbox{${\scriptstyle (a\rightarrow0)}$}}\,}
\def\upon #1/#2 {{\textstyle{#1\over #2}}}
\relax
\renewcommand{\thefootnote}{\fnsymbol{footnote}}

\sectioneq

\def\til#1{\centeron{\hbox{$#1$}}{\lower 2ex\hbox{$\char'176$}}}
\def\tild#1{\centeron{\hbox{$\,#1$}}{\lower 2.5ex\hbox{$\char'176$}}}
\def\sumtil{\centeron{\hbox{$\displaystyle\sum$}}{\lower
-1.5ex\hbox{$\widetilde{\phantom{xx}}$}}}

\begin{document} 

\begin{titlepage} 

\rightline{\vbox{\halign{&#\hfil\cr
% &ANL-HEP-CP-04-83\cr
&\today\cr}}} 
\vspace{0.25in} 

\begin{center} 

\centerline{The Nightmare Scenario and the Origin of the Standard Model.}
``We Got it Wrong ...How did we misread the signals? ... What to Do?'' 

\medskip

Alan R. White\footnote{arw@anl.gov }

\vskip 0.6cm

\centerline{Argonne National Laboratory}
\centerline{9700 South Cass, Il 60439, USA.}
\vspace{0.5cm}

\end{center}

\begin{abstract} 
It is thought that the emergence of the ``nightmare scenario'' at the LHC could 
be a serious crisis for particle physics that could require radical new concepts
and even a major paradigm change. A root cause may have been exaggeration of the significance of asymptotic freedom, leading to the historically profound mistake of formulating new short-distance extensions of the Standard Model while ignoring both serious infra-red problems and central elements of long-distance physics. In fact,
pursuit of the uniquely unitary Critical Pomeron 
leads to a possible gauge theory origin for the Standard Model that is both radical and paradigm changing, but also explains many mysteries. A bound-state S-Matrix embedded in a unique weak coupling massless SU(5) field theory emerges. The states and interactions of the Standard Model are enhanced, and the underlying SU(5) unification suppressed, by a wee parton divergence phenomenon involving wee gauge bosons coupled to S-Matrix massless fermion anomalies.  Confinement, chiral symmetry breaking, the parton model, electroweak symmetry breaking, dark matter, and neutrino masses, all appear to be present. Most significantly, perhaps,  there is a Higgs boson but, as seen experimentally at the LHC, there is no new short-distance physics. The only new physics is electroweak-scale QCD interactions due to color sextet quarks.

\end{abstract} 

\vspace{0.5in}

\renewcommand{\thefootnote}{\arabic{footnote}} \end{titlepage} 

\section{A Deep Crisis Requiring Radical New Concepts?}

In his overview talk\cite{DaGr} at Strings 2013, David Gross discussed the ``nightmare scenario'' in which the Standard Model Higgs boson is discovered at the LHC but no other
new short-distance physics, in particular no signal for SUSY, is seen. He called it the ``extreme pessimistic scenario'' but also said it was looking more and more likely and (if it is established) then, he acknowledged
\begin{center}
 {``We got it wrong.'' ``How did we misread the signals?'' ``What to do?''.}
\end{center}
He said that if it comes about definitively the field, and string theorists in particular, will suffer badly. He said that it will be essential for theorists who entered the field most recently to figure out where previous generations went wrong and also to determine what experimenters should now look for. 

In the following, I will argue that a root cause has been the exaggeration of the significance of the discovery of asymptotic freedom that has led to the historically profound mistake of trying to go forward by simply formulating new short-distance theories, supersymmetric or otherwise, while simultaneously ignoring both deep infra-red problems and fundamental long-distance physics.

In his recent ``Welcome'' speech\cite{NTur} at the Perimeter Institute, Neil Turok expressed similar concerns to those expressed by Gross. He said that
\begin{center}
``All the {\it \{beyond the Standard Model\}} theories have failed ... Theoretical physics is at a crossroads right now ... {\it \{there is\}} a very deep crisis.''
\end{center}
He argued that nature has turned out to be simpler than all the models - grand unified, super-symmetric, super-string, loop quantum gravity, etc, and that 
string theorists, especially, are now utterly confused - with no predictions at all. The models have failed, in his opinion, because they have no new, simplifying, underlying principle.  
They have complicated the physics by adding extra parameters, without introducing any simplifying concepts. 
 
The needed simplifying principle may simply be the fundamental long-distance requirement of full high-energy unitarity. The only known solution is the Critical Pomeron\cite{cri} that occurs uniquely (I have argued) in a bound-state S-Matrix that is embedded in 
QUD\footnote{QUD $\equiv$ Quantum Uno/Unification/Unique/Unitary/Underlying Dynamics}
- a very weak coupling (almost conformal) massless SU(5) field theory\cite{kw}, \cite{arw1}-\cite{arw8}, 
Remarkably, it seems that the S-Matrix states and interactions are those of the Standard Model. Consistency with the underlying SU(5) unification is achieved, not by a short-distance extension of the theory, but instead 
by the infra-red enhancement of the Standard Model interactions by a ``wee parton vacuum'' of anomalous wee gauge bosons coupled to S-Matrix massless fermion anomalies. 
There is a Higgs boson but there is no new short-distance physics
- just as is seen experimentally at the LHC,. The only new physics is electroweak-scale QCD interactions due to color sextet quarks (that could be discovered at the LHC) and there is no GUT scale. 

 The dynamical requirements of confinement, chiral symmetry breaking, the parton model and electroweak symmetry breaking all appear to be included in the bound-state S-Matrix, with dark matter and neutrino masses also present. Consequently, it seems that all the well-known Standard Model problems, including those which commonly motivate proposals for new short-distance physics, could be solved by the wee-parton anomaly solution of the long-distance problem of high-energy unitarity. Within the current theory paradigm, this is a radical proposition. Nevertheless, it may both simplify and unify the Standard Model, while dramatically changing expectations for new physics.

If it becomes accepted that supersymmetry and, by inference, string theory are not invoked by nature in the high-energy extension
and/or the unification of the forces of the Standard Model, then an enormous part of the research effort in theoretical high-energy physics, over the last few decades, will have been critically wasted. Also, as a consequence, vast resources will have been wasted on corresponding 
experimental searches. Perhaps, the worst consequence is for the future. As currently planned, the incredible power and potential of the LHC will  
be used at absurdly high luminosity, where only the existence of very rare events due to new short-distance physics (such as more exotic SUSY theories)
could possibly be discovered. Most likely, this physics does not exist and all that will be achieved is the extension of the nightmare scenario to even higher energy. 

In addition, if new electroweak scale long distance physics exists, as I am arguing for, the wasted opportunity will reach truly historic dimensions. This physics
could beautifully underly and unify the Standard Model but it is within a dynamical and philosophical framework that is very different to the current theory paradigm. 
Not surprisingly, therefore, it's unambiguous discovery would require the  
LHC accelerator and detectors to be operated in a very different manner.

Very unfortunately, also, the technical formalism of multi-regge theory that I use is unfamiliar to almost all of the current, short-distance educated, high-energy physics community. Moreover, an enormous amount of development is needed to provide a detailed implementation of my (so far only outlined) construction of the QUD bound-state S-Matrix.

\section{How Did it Happen?} 

It is extremely important to understand how the current situation has come about and, even more important, to determine what might be done to remedy it.
That Gross seems to have been the first leader in the field to publicly address the growing concerns is particularly remarkable since I will ascribe much of the ``wrong direction'' of the field to the unequivocal advocacy by him and his contemporaries of
a narrow (in the extreme) focus on short-distance physics, following the discovery of asymptotic freedom. Asymptotic freedom is a crucial field theory property that is responsible for strong interaction deep-inelastic scaling and, more generally, for the very existence of quantum field theories in the euclidean short distance region. However, the strong interaction nirvana that was anticipated to follow the discovery has not materialized.

The hope was raised that the strong interaction, and hence all interactions, could be calculated and describe experiments performed in a broad short distance region of phase-space. Indeed it might seem, at first sight, that this hope 
has been realized. However, it is at a level that is entirely phenomenological. By appealing to the parton model, physicists have become accustomed to focusing on 
short-distance physics and to thinking of quarks and gluons as physical particles. The calculation of perturbative QCD amplitudes has then allowed extensive ``beyond the Standard Model'' searches to be undertaken. Unfortunately, the validity of all of this thinking is completely undercut by the necessary introduction of a wide range of supplementary (parton model) assumptions and parameter dependent concepts that prohibit any pretense that the calculations involved are providing unique, fundamental, predictions. 

There is no derivation of the parton model and, consequently, there is no proof of the existence of parton distribution functions and certainly no possibility to calculate them. The current phenomenological application of the parton model involves the experimental determination of parton distributions from existing strong interaction data. The assumptions made are self-consistent only if it is assumed that QCD is well-understood in principle, even if not in detail, and so only new short-distance physics remains to be discovered. If there is new electroweak scale QCD physics to be discovered, of the kind that I will discuss, then the conventional parton model assumptions and procedure will clearly be inadequate.

On a deep level, the role and historical significance of asymptotic freedom has actually been that of a seductive siren enticing physicists to a shipwreck on the rocks of a belief that short distance physics can be a self-sufficient foundation for all future progress.  Focussing entirely on new short-distance theories, supersymmetric or otherwise, while putting aside serious theoretical infra-red problems and ignoring 
significant experimental and theoretical long-distance physics results, has been foolhardy at best and, at worst, may be a major cause of the nightmare scenario.  
Understanding both the origin of the parton model and it's coexistence with 
confinement is surely essential for long-term progress and demands top priority. It will be clear from the physics I describe that negelected long-distance physics could play a crucial role, with the origin of the parton model and confinement closely intertwined.

In Section 4, I will focus on an article written by Gross, entitled 
``Asymptotic freedom and the emergence of QCD'', that is an unrelenting advocacy of large transverse momentum physics. It was first published\cite{djg1} in 1992 
and then republished\cite{djg2}, without any changes, under the new title ``Twenty Five Years of Asymptotic Freedom'' in 1998. With added material and some editing, it was republished\cite{djg3} under the title
``Asymptotic freedom and QCD: A historical perspective'' and finally, it was again republished\cite{djg4}, with further modifications, as his Nobel lecture ``The discovery of asymptotic freedom and the emergence of QCD''. Perhaps unfairly, I will focus on the initial article as representative of the ``We got it wrong'' element of the history of asymptotic freedom. 

The aim of the article was to provide both an outline of the discovery of asymptotic freedom and a broad historical description of the background research environment. Regrettably, in addition to appropriately emphasizing 
the significance of asymptotic freedom, there are also a number of inaccurate, misleading, and even wrong, arguments provided to make the case for a research program (theoretical and experimental) based solely on large transverse momentum physics,
to the exclusion of other more complex, more difficult, and ultimately I believe, more fundamental directions.

A basic (and surely revealing) problem is apparent in the opening paragraph. In the second sentence, Gross says ``history is written by the victorious''. He saw his research as part of a battle between quantum field theory and S-Matrix theory and, as he describes it, the battle was intense, long-lasting, and at times acrimonius. It was perceived as a fight for the very survival of quantum field theory and, indeed, it is possible that today's nightmare scenario has come to pass as a direct, and profound, consequence of the all or nothing level of the fight. In a battle, all the spoils go to the victor and, apparently, to Gross and many others this meant that the discovery of asymptotic freedom had swept field theory to victory and vanquished all research programs based on, or even having any connection to, S-Matrix theory. The omnipotence of field theory was established! 

The most serious problem threatening the existence of field theories had always been the uncontrollable (wildly divergent) behavior of the perturbation expansion in the large momentum region. In an asymptotically free theory this problem is removed and so does not prevent the existence of such theories. However, all the problems are transferred to the infra-red region and so, as has to be strongly emphasized, a jump to the existence of a full non-perturbative quantum field theory containing, in particular, correlation functions with all the conventionally assumed properties, is enormous and much, much, more (beyond asymptotic freedom) would be needed. The problems involved can not be addressed within the framework of short-distance field theory and, in fact, they are so serious that no four-dimensional field theory has ever been shown to exist and, most likely I would argue, never will be.

Gross also talks disparagingly about being taught, as a graduate student, that {\it Field Theory = Feynman Rules} and argues that this attitude was diverting attention from more important non-perturbative issues. Yet still today, in reality, the Feynman integrals are the only well defined formulation of a non-abelian theory such as QCD. What non-perturbative quantities they represent (if any) has still to be determined. The Feynman path integral, which is universally assumed to be the desired non-perturbative formulation, is deceptively alluring and is surely a very powerful formal tool. Unfortunately, because of the four-dimensional infinite volume divergence and the undefined nature of the function space implied by the Gribov copy problem, it is very unlikely to actually exist. Indeed, there is a ``Catch-22'' element to this issue. To prove the existence of the infinite volume 
limit requires an effective infra-red lagrangian without massless particles, but such a lagrangian would have to be derived from the (previously defined) infinite volume integral. It is important to note that the extensive, and largely successful, physical applications of lattice gauge theory always use finite volume approximations to the path-integral which, in effect,
approximate finite momentum contributions to perturbative feynman 
integrals and which, in principle at least, could be Borel summable.

\section{The Way Forward?}

It would surely be much more satisfying, ultimately, to understand that nature has not been fickle in failing to take advantage of the beautiful elegance of supersymmetric field theories. Rather it may have not done so simply because such theories retain infra-red problems that actually 
prevent them from providing the basic necessity for a particle theory, namely a massive spectrum of physical states with a unitary S-Matrix. 
As I discuss further in Section 5, the perturbative infra-red problems of an asymptotically-free gauge theory are even more serious than the ultra-violet
problems that have been solved. A priori, an infinite number of  
vacuum condensate parameters is introduced by the wild divergence of the perturbation series and there is no evidence that the conventional formulation
of confinement, even if it could be proved, would be sufficient to allow a physical S-Matrix to exist. It can not be assumed that serious
 infra-red problems of this kind simply take care of themselves and that physicists need only calculate ``safe, infra-red finite'' short-distance cross-sections. 

I will argue that the requirement of a consistent, unitary, particle S-Matrix is so strong that it is satisfied only in very specific circumstances in which the dynamics actually excludes the existence of off-shell correlation functions.
To arrive at this conclusion it is necessary to accept the seriousness of all of the infra-red problems that are inherent in the formulation of quantum field theories containing massless fields. It is also essential to incorporate the
theoretical results and understanding  of high-energy unitarity acquired in the context of forward physics experimental results that have motivated my search for, and discovery of, the Critical Pomeron in a gauge theory. As I discuss in the next Section, in his article, Gross foolishly derides these experimental results as having provided no insight into the dynamics of the strong interaction. 
If I am right, they provide the ultimate key to the origin of the Standard Model.

Multi-regge S-Matrix theory provides a vehicle for handling elaborate multiparticle Feynman diagrams in the generalized infinite momentum kinematics
of multi-regge limits. Remarkably\footnote{The multi-regge formalism may be alone in allowing the study, in four dimensions, of gauge theory bound-state amplitudes.}, perhaps, the emergence of full bound-state amplitudes via infra-red divergences can be studied. As I will outline in later Sections, I have used this formalism to argue that the existence of a bound-state unitary S-matrix with the desired forward physics properties, including
unitary Critical Pomeron high-energy behavior, could be a very special property of
QUD - the unique SU(5) massless field theory that I discussed in the last Section. The infinite-momentum dynamics involves massless fermions in a fundamental manner.
Combinations of infra-red divergent anomalous gauge bosons coupled to massless fermion chirality transitions, produce 
``universal wee parton'' gauge bosons in all S-Matrix multi-regge amplitudes. It is  deeply significant that the
anomalies responsible for the chirality transitions are only present in on-shell infinite-momentum amplitudes so that, as a result, the
formation of bound-states and S-Matrix interactions is inexorably linked.
The outcome is a very special version of QCD and ultimately, I believe, the origin of the Standard Model.   

It will be truly, truly, ironic if the physics that I describe does actually provide the way forward from the nightmare scenario. 
Most likely, the S-Matrix is the only well-defined non-perturbative element of QUD and I believe that this is what the Standard Model is reproducing. Unification of the interactions is achieved, not by a short-distance extension of the theory, but instead by the underlying wee parton 
structure present in the S-Matrix as a consequence of long-distance anomaly dynamics. Moreover, the very different dynamical role played by fermions (via anomalies) and gauge bosons (via infra-red divergences) makes it apparent that this dynamics could not be present in a supersymmetric theory. Indeed, it may be that the field theory is only  well-defined 
in the short-distance region where asymptotic freedom is operative and that, outside of this region, only the particle S-Matrix is well-defined. This has very important scientific and philosophical implications, as we discuss later. Certainly, the
misleading and counterproductive nature of Gross's arguments (described further in the next Section) that short-distance physics should be the focus of experiments is clear and the resulting historical misdirection glaring. 

The simplification provided by QUD, as well as the radical philosophical and conceptual changes implied, surely goes in the direction anticipated by Turok. Obviously,
the scientific and aesthetic importance of an 
underlying and unifying non-supersymmetric massless field theory for the Standard Model 
can not be exaggerated. QUD is self-contained and is either entirely right, or simply wrong! The only new physics is a high mass color sextet quark sector of the strong
interaction that gives electroweak symmetry breaking and dark matter. Moreover, many other fundamental puzzles appear to be explained.

\section{Only Short-Distance Physics Matters!}

In this Section I will discuss David Gross's article in some detail. I will list a few quotes (in the order in which they appear) that I think best illustrate the attitudes and priorities that (I believe) have contributed to the occurrence of the nightmare scenario.

\begin{enumerate}

\item{{\it ``Field theory was in disgrace; S-Matrix theory was in full bloom. ..
 A powerful dogma emerged - that field theory was fundamentally wrong,
 especially in its application to the strong interactions ...
 was to be replaced by S-matrix theory; a theory based on general principles, such as unitarity and analyticity, ...''}

That a battle is to be fought for the survival of field theory is crystal clear. As far as possible, after victory, all elements of S-Matrix theory, including much very powerful large distance formalism, will be dismissed as irrelevant dogma.}

\item{{\it ``The basic dynamical idea was that
there was a unique S-Matrix that obeyed these principles. ... This is
of course false. We now know that there are an infinite number of consistent S-Matrices that satisfy all the sacred principles. One can take any non-Abelian gauge theory, with any gauge group, and many sets of fermions (as long as there are not too many to destroy asymptotic freedom).''}

The claim of non-uniqueness of the unitary S-Matrix is unbelievably misleading and, most likely, completely wrong. As I have already emphasized, gauge theory S-Matrices can only be calculated perturbatively and the  D=4 expansions for every field 
(and string) theory are wildly divergent and, almostly certainly, can not be summed. There is certainly no non-perturbative
formulation of any theory that can derive S-Matrix 
amplitudes - let alone discuss unitarity.

A need to worry about the existence of a unitary S-Matrix would surely have
been a severe constraint on the wild proliferation of short-distance
field theories (discussed in the next Section) that have been formulated without any concern for potential infra-red problems. The claimed non-uniqueness of the S-Matrix, by Gross, was intended to discredit the basic ingredient of the bootstrap program that was the early dynamical centerpiece of S-Matrix theory. However, although the bootstrap
program has long since disappeared, uniqueness of the unitary S-Matrix may deeply intertwine with the origin of the Standard Model, as I have already suggested and will return to later.}

\item{{\it ``theorists and experimentalists reinforced each other’s conviction that the
secret of the strong interactions lay in the high-energy behavior of scattering amplitudes at low momentum transfer. ... prompted by the
regularities that were discovered at low momentum transfer, theorists developed an explanation based on the theory of Regge poles. This was the only strong interaction dynamics that was understood, for which there was a real theory. Therefore theorists concluded that Regge behavior
must be very important and forward scattering experiments were deemed to be the major tool of discovery.''}

This is, perhaps, the strongest contributor in the category of ``How did we misread the signals?'' It is both ignorant and scandalously arrogant.
All the beautiful experimental results on the regge behavior of forward cross-sections, most deeply the isolated regge
pole nature of pomeron exchange, are dismissed as resulting from theorists and experimentalists having nothing else to talk to each other about. In addition, 
fundamental theoretical results based on dispersion/regge theory analysis of unitarity are to be thrown away as irrelevant. 

In fact, as a crucial outcome of the disdained experiments, it was definitively established that, in first approximation, the pomeron 
is a factorizing single Regge pole with no resonances on the trajectory.
Because later, higher energy, experiments did not measure a variety of scattering processes, and because the forward region was very inadequately explored, it became possible for this inconvenient truth to be universally, and persistently, ignored (by BFKL enthusiasts in particular). However, recent LHC results have now beautifully
confirmed the existence of the lower-energy diffraction peak at the highest energy. 

\begin{center}
\epsfxsize=5.5in \epsffile{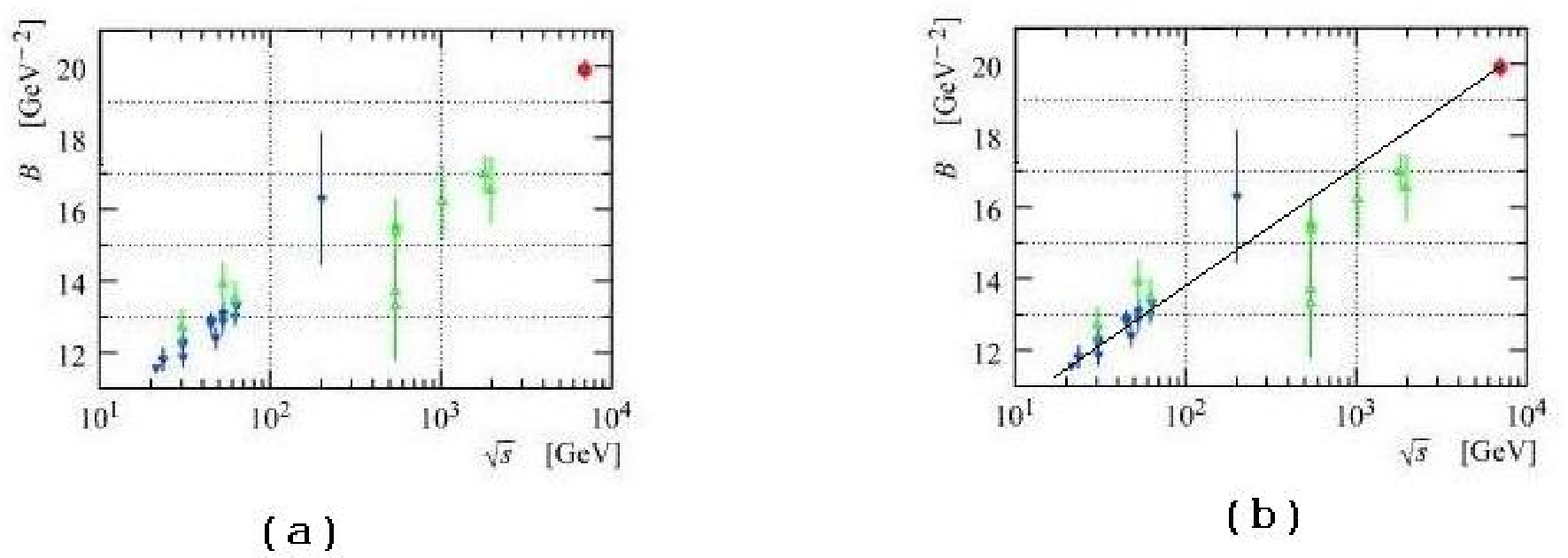}

Figure 1. TOTEM Measurement of the Forward Elastic Slope, (a) Data, (b) The Linear Extrapolation
\end{center}
The plot of measurements of the forward elastic slope shown in Fig.~1(a) is taken from a recent TOTEM paper\cite{TOT}. It illustrates the point perfectly. 
The TOTEM measurement (red dot) is stunningly accurate and, as is shown by the hand-drawn line in Fig.~1(b), very clearly lies on a linear extrapolation of the lower energy results. The linear energy dependence of the slope is, of course, a direct property of a Regge pole amplitude.
Also evident in Fig.~1(a) is the inaccuracy of the intermediate energy results, reflecting the lack of priority given to the experiments involved. Without the TOTEM result, the freedom of invention for forward physics theorists (at the highest energies) would be largely unfettered! 

As I will return to later, an approximate regge pole pomeron is deeply important for the existence of the parton model in QCD. It is very fortunate that a small minority of experimenters have
continued to pursue the forward physics that has been pushed out of the mainstream by the short-distance mania that Gross and company have incited! 

It will be very apparent by the end of this article that reproducing the forward physics seen in experiments is the key to understanding the present mysteries of QCD. All that can be said, apparently, from Gross's lofty viewpoint is that}
 
 \item{{\it ``... the explanation of Regge behavior remains an interesting, unsolved and complicated problem for QCD.''}

All of the elaborate, and extensive, results on the reggeization of both gauge bosons and fermions in non-abelian 
gauge theories are considered to be irrelevant. Not, surprisingly I will also return to this subject later.}

\item{{\it ``It was not at all realized by theorists that the secret of hadronic dynamics could be revealed by experiments at large momentum transfer that probed the short distance structure of hadrons.''}}

\item{{\it ``Only much later, after the impact of the deep inelastic scattering experiments that had been ridiculed by many as unpromising, was it understood that the most informative experiments were those at large momentum transfers that probe short or light-like distances.''}

The almost universal, uncritical and unqualified, acceptance of these last statements has probably been as damaging as any other consequence of the discovery of asymptotic freedom. Taken literally,
they are\label{key} wildly untrue. With no derivation of the parton model and 
the associated parton distribution functions, 
no derivation of confinement, and no derivation of ``non-perturbative''
contributions, initially precise short-distance calculations are submerged
in a phenomenological sea. As a result, the only unambiguous information produced by short-distance experiments is that the strong interaction can be described by a non-abelian gauge theory. The details of which gauge group, which fermions, the mass spectrum, all come from long-distance experiments and would be very difficult to derive directly from the comparison of short distance calculations with strong interaction experiments.}

\item{{\it ``Thus the discovery of asymptotic freedom greatly reassured
one of the consistency of four-dimensional quantum field theory. ...
We are very close to having a rigorous
mathematical proof of the existence of asymptotically free
gauge theories in four dimensions – at
least when placed into a finite box to tame the infrared dynamics that produces confinement. As far as we know, QCD by itself is a totally consistent theory at all energies.''}

The introduction of the finite (euclidean) box is viewed as little more than a technical
convenience. Apparently, all that remains is for confinement, which is not understood at all, to be elaborated sufficiently to allow the box to be taken away. As I will discuss more in the next Section, this is an incredible (head in the sand) attitude
towards all that would have to be done to allow consistency of QCD at all energies to be claimed.}
\end{enumerate}
Some (well-known) quotes from Feynman seem particularly relevant, with the second perhaps addressing the origin of the nightmare scenario most directly.

{\it ``The first principle is that you must not fool yourself and you are the easiest person to fool.''

``It is in the admission of ignorance and the admission of uncertainty that there is a hope for the continuous motion of human beings in some direction that doesn't get confined, permanently blocked, as it has so many times before in various periods in the history of man.'' }
 
\section{The Infra-red Mysteries of QCD Can be Shelved!}

While the discovery of asymptotic freedom has clearly led to the idea that the focus of physics should be on short-distance processes, it has also produced 
an even greater acceptance of the idea that all the rest of physics is 
both unfathomable and unproductive to study. At the same time,
paradoxically but very conveniently, it has  become widely believed that QCD is sufficiently well understood that the following list of problems can be ``shelved'' while the 
march towards the discovery of new short distance physics proceeds.

\begin{enumerate}
\item{There is no proof, and not even an argument, that the euclidean functional integral exists, and so there is no known way to define QCD as a quantum field theory, outside of perturbation theory.}
\item{The infra-red renormalons determine that the perturbation expansion is wildly divergent and is not Borel summable, except when the maximum number of massless quarks allowed by asymptotic freedom is present. In this special case, 
which is at the core of my discussion later in the paper, there is an infra-red fixed-point which eliminates the renormalons. As a result, the theory is almost conformal and so lies in the
``conformal window''. This is generally thought to imply that it can not have a massive particle spectrum. However, it actually could have an S-Matrix with a spectrum including massive particles{\it - provided there are no corresponding off-shell amplitudes.}}
\item{There is no derivation of the parton model. Assuming, nevertheless, that it exists, the factorization 
theorems needed for self-consistency apply only to deep-inelastic leading-order perturbation
theory. They do not apply, at all, to hadron scattering.}
\item{There is no understanding of the confinement of color and
chiral symmetry breaking. Most importantly, there is no understanding of how these properties can coexist with the parton model. }
\item{There is no derivation of parton distribution functions and so, necessarily, their formulation and application is entirely phenomenological.
Consequently, as I have discussed\cite{arw6} in previous papers the presence or absence of new, electroweak scale, physics in strong interaction cross-sections can be a matter of choice! Morover, infra-red finiteness has apparently become the
only criterion that must be satisfied for physical cross-sections to be derived from QCD.}
\end{enumerate} 

The above QCD problems are universally acknowledged as unresolved.
I would also add two more that have emerged from the unfathomable region of small
transverse momentum physics and that also pose a major challenge to the conventional understanding of QCD. They are closely related.
\begin{itemize}
\item{There are no glueballs seen experimentally, even though they are widely believed to be a general consequence of QCD. This suggests that the
nature of confinement is more selective than conventionally thought. }  
\item{As discussed in the previous Section, the forward scattering experiments disdained by Gross established that, in first approximation, the pomeron 
is a factorizing single Regge pole. Recent LHC results have beautifully
confirmed this result. It is a complete contradiction with the conventional description of near forward  physics via the BFKL pomeron.}
\end{itemize} 

It seems to have become accepted that this long list of infra-red problems, to which should be added the most important of general principles - non-perturbative unitarity, are irrelevant in the search for a physical theory.
The paradigm has become that theories should first be discovered 
by their ability to fit experimental facts. If necessary, consistency can be looked for afterwards. Indeed, the conventional wisdom has become that, because of it's success in describing experiments, the Standard Model must be a well-defined quantum field theory that, necessarily, must have all of the desired properties of a full, non-perturbative, theory\footnote{\openup-1\jot{Even 
though it has long been acknowledged that a mathematical effort of unimaginable, herculean, magnitude would be required to prove this\cite{jw}.}} - with a massive unitary S-Matrix automatically included. 

Amazingly, as I believe it will be seen in retrospect, it is also commonly
thought that, while
the Standard Model fits the experimental data, it is just one 
of an infinity of renormalizable field theories that nature could have chosen. From this perspective, it has been assumed that progress beyond the Standard Model will be determined by new experimental
phenomena that can be similarly fitted by an enlarged field theory. 
Since (it has also been thought that) long distance physics is sufficiently well understood via QCD, the new phenomena should appear at short distance. Given Gross's assurance, 
theorists that have speculated about new physics that might appear have  
not been fettered by any fear that the long-distance physics demand for
unitarity of the physical S-matrix could make any significant selection
amongst candidate theories.

As I hope will be crystal clear by the end of this article, I am arguing that progress requires understanding both why QCD is 
the only gauge theory with all the necessary requisites to describe a unitary strong interaction and how it has to be modified to make it a fully consistent theory.

Until the advent of the nightmare scenario, the general belief has been that there will be
a unifying quantum field theory that includes the Standard Model
and which (if quantum gravity is to be 
included) will embed in an (ever more elusive) string theory.
It is rarely acknowledged that the extra (far from trivial) assumption has to be made that unification is possible without conflict between the intrinsic non-perturbative applicability of QCD (involving 
confinement) and the perturbative applicability of the electroweak sector. 
Although there is no explicit understanding of how it could happen, it is thought that a transition from perturbative physics to non-perturbative confinement physics can simply be a consequence of the evolution of couplings with the scale involved.

As I noted in the previous Section, Gross considered that regge behavior is merely {\it ``an interesting, unsolved, and complicated problem for QCD''}.
This has been a widely held view that, I would argue, could not be further from the truth. The regge region is where the 
connection between perturbative and ``non-perturbative'' physics should be explicitly evident, since a mixture of small and large momenta is involved.
Moreover, multi-regge theory provides the central framework for our 
discussion of the existence of a unitary S-Matrix in the following. Regge-region (reggeon) unitarity is deeply related to all the fundamental problems in the formulation of QCD and is central in our construction of a fully unitary gauge theory. 

In fact, the nightmare scenario implies there is simply no experimental support for the viewpoint, long held by many, that the more difficult dynamical problems in QCD can be put aside and replaced by
the guiding principle/paradigm that progress will come via inspired guesses for missing short-distance physics, combined with experimental 
verification via predicted rare processes. It should also be emphasized that there is no historical precedent supporting such a viewpoint.

As I have already implied, my solution to this situation satisfies, perhaps, Turok's requirement that new simplifying principles be involved. Insistence on high-energy unitarity of the S-Matrix, as a principle, may actually uncover the desired extension of the Standard Model. Moreover, an extensive revision, both practical and philosophical, of the current theory paradigm is necessarily involved. 

\section{Why is S-Matrix Unitarity a Deep Requirement?}

During the barren years for quantum field theory, that produced the development of S-Matrix theory (before the discovery of asymptotic freedom!), basic questions concerning the necessary and sufficient elements of a quantum theory of particle physics were intensely discussed. The conclusion was that the minimum requirement is a unitary S-Matrix that describes the scattering of particles with a massive spectrum. The additional superstructure of off-shell Green's function that a field theory provides (at least in perturbation theory) is both unphysically detectable and unnecessary - unless, as is often discussed nowadays, a short-distance unification with quantum gravity is desired.

Paradoxically, it is the unnecessary superstructure of off-shell
short-distance amplitudes (in an infra-red cut-off field theory) that asymptotic freedom has shown the existence of. In fact, maximizing the physical consequences of asymptotic freedom requires, not only the simultaneous existence of a physical spectrum of particles and an infra-red finite S-Matrix, but also some form of parton model that allows asymptotic physical amplitudes to be expressed in terms of elementary field theory amplitudes,

The possibility that a field theory S-Matrix could exist without corresponding off-shell amplitudes has not been commonly envisaged and so cavalier assumptions that confinement will provide wave-functions coupling physical states to
off-shell amplitudes are often made. Surprisingly, perhaps, I will argue that 
the existence of a parton model is a very special property that is connected with 
the origin of both the Reggeon Field Theory 
Critical Pomeron  and the Standard Model and is in conflict with 
the existence of off-shell physical amplitudes.

In the midst of the various publications of David Gross's article, I published a 
review\cite{arw9} 
describing some of the major results of S-Matrix theory and included a discussion of the Critical Pomeron and how it might connect to a non-abelian gauge theory. Here are some quotes suggesting that the disdained pomeron physics might point the way towards the origin of a unitary S-Matrix.

\begin{enumerate}
\item{ {\it ``... the Critical Pomeron is the summit of abstract S-Matrix Theory.
It satisfies all known unitarity constraints ...
provides a uniquely attractive possibility for ... an S-Matrix 
satisfying the maximum strength postulate.''}

\noindent To potentially match with large transverse momentum gauge theory amplitudes, a regge pole pomeron must have unit intercept, i.e. it must satisfy the maximum strength postulate.}
\item{\it ``If the Critical Pomeron is the only
high-energy solution of unitarity that can match with asymptotic
freedom then perhaps there is a uniqueness property for the
strong-interaction S-Matrix ...''}
\item{{\it But, why should Critical Pomeron asymptotic behavior be unique? Why ... should the pomeron be only a single regge 
pole plus multipomeron cuts?} 

 It was introduced just because this is exactly what forward scattering experiments (that Gross said gave no information) tell us is the case!}
\item{\it  A 
single regge pole ... uniquely has the factorization 
properties needed ... a universal wee-parton  distribution 
in hadrons. ... the maximal 
applicability of short-distance perturbation theory. ... may 
well be essential to produce a completely finite (and unitary) S-Matrix.}
\item {\it Could the full S-Matrix including 
the electroweak interaction be unique?}
\item {\it ... the uniqueness of the S-Matrix determines the underlying 
gauge theory, before ... gravity, ... would be 
strongly counter to today's prevailing philosophy.''}
\end{enumerate}

The existence of a short-distance field theory may be essential, not only for  
the large momentum finiteness of the S-Matrix, but also, as discussed in \cite{arw9}, 
for local analyticity properties. 
In later Sections I will outline how the unique SU(5) massless theory, QUD, appears. In the final Section, I will discuss why, in line with the above quotes, the properties that select this particular field theory may be essential for the existence of a physical S-Matrix. 

As I have already noted in Section 4, it is currently accepted, almost without question, that ``non-perturbative'' QCD 
and all similar
unbroken non-abelian gauge theories should be
well-defined by the euclidean path integral. This is taken to imply that
there must be a physical S-Matrix and that, moreover, the physical states appear as
intermediate states in off-shell Green's functions (derived from the path 
integral) of appropriate operators. Although there is no evidence to support
this hypothesis, the considerable 
phenomenological success of ``non-perturbative QCD'' formalisms, particularly
lattice QCD, implies there must be some approximate truth in the assumptions.
Nevertheless, at the level where we are concerned with whether a particular theory is uniquely chosen by nature, it is important to emphasize that approximations are being made and that there are significant assumptions involved. (As I have emphasized earlier, lattice applications always use a finite volume approximation
- that can, in principle, be related back to feynman diagram contributions.) 

That the S-Matrix can be obtained from ``non-perturbative'' off-shell Green's functions 
does not appear to be essential for any of it's basic properties. The
global analyticity domains that are normally thought to be a consequence of an 
off-shell field theory probably follow from the construction of physical high-energy 
amplitudes via the perturbation expansion. In fact, when the fields are massless and 
bound states related to infra-red anomalies are involved there is probably no 
general reason to expect a connection between Green's functions and the S-Matrix. 

In general, there is not even a formal
property of a non-abelian gauge theory path integral which implies that a 
unitary, bound-state, S-Matrix can be derived via Green's functions.
Even worse, as I noted earlier, because of infra-red problems, the path integral itself is, most likely,
not well-defined. Since there are no ``non-perturbative'' methods for constructing gauge theory S-Matrix amplitudes that do not, effectively, appeal to the formal euclidean functional integral, 
to seriously discuss whether a unitary S-Matrix exists in a general gauge theory 
is a highly non-trivial problem.

\section{The Supercritical Pomeron and QCD}

Although motivated directly by the forward scattering experiments, it was not clear what the deep significance of a regge pole critical pomeron might be. Since a Reggeon Field Theory renormalization group fixed-point is involved
it might seem that there could be a link with the asymptotic freedom
calculations. However, the pomeron was thought to involve complicated composite degrees of freedom that have no simple connection to the underlying gauge fields. Nevertheless, 
there was a mystery. While the regge pole pomeron that appears in experiments does not appear perturbatively in any gauge theory, all the elementary fields 
are associated with reggeizing particles. Naively, this might suggest that
(in contradiction with confinement) the pomeron should be related to a
non-abelian gauge boson and baryons should be gauge theory fermions. I will describe why, remarkably, this suggestion might be much closer to reality than conventional expectations would imply.

That the Critical Pomeron is selective in it's association with an underlying field theory is seen directly via the approach from the supercritical side\cite{arw10}. The even signature of the pomeron requires a pure imaginary triple pomeron vertex that makes the effective action non-hermitian. Consequently, there is no supercritical minimum and only supercritical stationary points exist. As a result, a graphical expansion containing a pomeron field condensate is not straightforwardly obtained. 
Instead, a ``supercritical pomeron condensate'' is introduced as a zero transverse momentum component of the scattering states - via multi-regge theory.
There is then a supercritical regge pole pomeron, together 
with an exchange degenerate vector regge pole.
There are also singular vector exchange interactions due to the wee parton ``pomeron condensate''. In effect, the rapidity divergences of the bare pomeron 
are replaced by transverse momentum divergences due to the vector particle. 

The vector reggeon immediately suggests that the supercritical pomeron might be found by starting from the reggeon diagrams of
color superconducting QCD\footnote{Color superconducting QCD reggeon diagrams potentially resolve the quantization problems of Gribov copies and Gauss' law, via zero momentum longitudinal gluons.}. However, a smooth (asymptotically free) connection to unbroken QCD is realistically possible only if color sextet quarks produce electroweak symmetry breaking. A further constraint is that the pomeron condensate has to originate from gluon infra-red divergences that, to avoid the exponentiation of reggeization, must couple to anomalies that appear only when massless quarks are present. 

Because the contribution of arbitrarily high order feynman diagrams is included, the analysis of  multiparticle multi-reggeon diagrams that I have outlined in previous papers, systematically including the anomalies, is potentially
the most extensive study of the infra-red divergences of QCD feynman diagrams that has yet been formulated. I consider multiparticle amplitudes within which
bound-state scattering amplitudes can occur. Triangle anomaly diagrams occur in reggeon interactions that connect different (rapidity and transverse momentum)
reggeon channels. Zero mass quarks generate anomaly poles in the triangle diagrams via chirality transitions that are zero momentum Dirac sea shifts of
positive to negative (or vice versa) zero energies. As a result, chirality
transitions determine how reggeon states in different channels couple, but do not contribute to the dynamical formation of individual reggeons.

I consider color superconducting QCD in the di-triple-regge region, where gluon divergences coupled to anomaly poles appear simultaneously in bound-state and interaction channels. An analysis involving reggeon interaction kernels 
shows that almost all infra-red divergences are exponentiated by reggeization, except for an overall divergence, which  
is subtracted to define physical amplitudes. A color confining ``parton model'' appears
in which
``anomalous wee gluons'', produced by multi-reggeon generalizations of the well-known anomaly current, provide vacuum-like universal wee partons. Anomaly poles coupled to the wee gluons both produce chiral Goldstone particle 
poles\footnote{Via the triangle diagram, a Goldstone anomaly pole is both a simple quark/antiquark state, with one of the pair having zero momentum and zero negative energy or, alternatively, it is a
physical reggeon state containing physical quark/antiquark reggeons plus anomalous wee gluons.} and also couple the wee gluons in distinct reggeon channels. 

The color compensation by anomalous wee gluons  beautifully resolves the mystery of the connection between a regge pole pomeron and gauge theory reggeization.
$SU(2)$ anomalous gluons have $\tau=-1 = -C$ and so a  $\tau=+1$ supercritical 
pomeron appears composed of an SU(2) singlet massive gauge boson reggeon plus anomalous wee gluons. It is
exchange degenerate with a massive gluon reggeon, just as in 
supercritical RFT. The anomalous, $C= -1$, color charge parity of the pomeron is directly linked to the chiral symmetry breaking nature of the anomaly pole bound-states.

If all the quarks are massless, there is an infra-red fixed-point and the anomalous wee gluon divergence remains as arbitrarily higher-order reggeon diagrams are included. It effectively produces the desired supercritical pomeron condensate and correctly reproduces the supercritical pomeron interactions. Potentially,
the final result is a confining, chiral symmetry breaking, spectrum generated in superconducting massless quark QCD.

Very importantly, the condensate remains after the restoration of the full
SU(3) color symmetry and so it is present in both the physical pomeron and the physical bound-states of the critical theory. In effect, the supercritical phase involves color symmetry breaking of the condensate as well the production of a massive vector reggeon. 

\section{The Critical Pomeron in Massless QCD$_S$}

The need for a massless quark infra-red fixed point determines 
that the Critical Pomeron can occur in QCD only when the maximum number of 
massless quarks is present. The only ``semi-realistic'' possibility, that we already arrived at by wanting to start from superconducting QCD, is massless QCD$_S$ in which there are
six massless triplet quarks and two massless sextet quarks.
Because of the infra-red fixed point there are no infra-red renormalons and the perturbation expansion is much less divergent - without 
the array of multi-gluon condensates normally produced. Consequently the possibility that perturbation theory (sums) can produce meaningful results is much improved. ``Non-perturbative''
physics could be provided, in principle at least, by topological (multi-instanton) contributions. The theory is ``almost conformal'', with the infra-red fixed-point implying that if off-shell correlation functions exist (to
which a renormalization group scaling transformation can be applied) they can have only scale-invariant intermediate states. Therefore, if there are massive physical states in massless QCD$_S$, off-shell correlation functions containing these states can not exist.

At first sight, there are many experimentally desirable features of the states that appear. 
\begin{itemize} \openup-1\jot{
\item{Potential bound-states are triplet and sextet (pseudoscalar) mesons, together with triplet and sextet baryons.}
\item{ There are no hybrid sextet/triplet states.} 
\item{There are no glueballs. Both the BFKL pomeron and the odderon are absent because they do not couple to states containing anomaly poles.}
\item{The Critical Pomeron is a regge pole, plus a triple pomeron interaction.}
\item{Sextet anomaly color factors imply larger (electroweak scale) masses for sextet states.} 
\item{ Wee gluon color factors imply large high-energy pomeron cross-sections for sextet states.}}
\end{itemize}
These features would be realized if a QCD$_S$ bound-state S-Matrix exists. However, because of the large array of chiral symmetries, there would necessarily be a large multiplicity of
massless Goldstone bosons that would create, probably insuperable, infra-red problems
for the existence of such an S-Matrix. Moreover, that 
the Critical Pomeron appears only in massless QCD$_S$ makes it's appearance in a massive hadron theory seem very unlikely, if not impossible! Fortunately, this conflict
is resolved by the embedding of QCD$_S$ in QUD, as described in the next Section. 

A priori, the critical behavior involves zero momentum quark/antiquark chirality transitions 
(Dirac sea shifts), due to initial SU(2) anomalous wee gluons, becoming random dynamical fluctuations associated with general anomalous wee gluons  within the full SU(3) group. Note that, since the chirality-transition anomalies are a multi-regge S-Matrix phenomenon that can not produce off-shell correlation functions, reproducing the same physics at finite momentum would surely be very challenging. Since the wee gluon configurations all couple to multi-fermion instanton interactions we could perhaps, very loosely and for conceptual purposes only, think of the instanton interactions as responsible for the wee parton ``vacuum'' within which perturbative interactions operate. In practise, of course, this would be impossible to demonstrate directly.

\section{The Critical Pomeron $~\longleftrightarrow~$ QUD}

A remarkable result emerges when we consider combining the electroweak interaction with the Critical Pomeron. We discover a unique theory\cite{kw,arw1} that is again massless, asymptotically free, and saturated with fermions, but has left-handed couplings to all fermions. Requiring asymptotic freedom and no anomaly, massless QCD$_S$ and the electroweak interaction embed
uniquely  in QUD, i.e. SU(5) gauge theory with left-handed massless fermions in the 
$5 \oplus 15 \oplus40 \oplus 45^*$ representation. 

Under $ SU(3)\otimes SU(2)\otimes
U(1)$
\openup-0.7\jot{ 
{\footnotesize $$ 
5=(3,1,-\frac{1}{3}))
+(1,2,\frac{1}{2})~,~~~~~ 15=(1,3,1)+
(3,2,\frac{1}{6}) + (6,1,-\frac{2}{3})~,
$$
$$
40=(1,2,-\frac{3}{2})
+(3,2,\frac{1}{6})+
(3^*,1,-\frac{2}{3})+(3^*,3,-\frac{2}{3}) + 
(6^*,2,\frac{1}{6})+(8,1,1)~,
$$
$$
45^*=(1,2,-\frac{1}{2})+(3^*,1,\frac{1}{3})
+(3^*,3,\frac{1}{3})+(3,1,-\frac{4}{3})+(3,2,\frac{7}{6}))+
 (6,1,\frac{1}{3}) +(8,2,-\frac{1}{2})
 $$}}
 
 There is an infra-red fixed-point but now there are no exact chiral symmetries and so all bound-states should acquire masses.
In fact, QUD has all the additional structure needed to generate, via massless fermion anomaly dynamics, a bound-state S-Matrix that reproduces the full Standard Model.
The only additional elements are a dark matter sector and neutrino masses, both of which are extremely welcome.  

There are three ``generations'' of both elementary leptons and elementary triplet quarks, and the theory is vector-like with respect to SU(3)xU(1)$_{em}$. SU(2)xU(1) is
not quite right but, in the S-Matrix constructed\cite{arw1,arw2,arw6} via multi-regge theory, all elementary fermions are confined and only Standard Model interactions and states emerge. The color sextet sector provides ``sextet pions'' that produce electroweak symmetry breaking and sextet baryons - with the sextet neutron and antineutron providing stable, massive, dark matter particles. There is also a color octet sector 
that is responsible, via large $k_{\perp}$ anomalies, for the generation structure of 
the physical states.
 
\subsection{Infra-Red Analysis of QUD Reggeon Diagrams}
 
In QCD$_S$ the chirality transitions do not conflict with the vector gauge symmetry. In QUD they break the {\bf non-vector} part of the gauge symmetry. The QUD reggeon diagrams are initially well-defined, if all reggeons are given masses.
24 and 5$\oplus$5$^*$ scalars give initial masses to all fermions and their decoupling
 leaves chirality 
 transitions that break SU(5) to SU(3)$_C\otimes$U(1)$_{em}~$ {\bf in reggeon anomaly vertices only.} Using 
 5$\oplus$5$^*$ scalars for the gauge bosons gives a smooth massless limit (via complementarity). Their successive
 decoupling gives reggeon
 global symmetries 
 \newline $~$
 \newline
 \centerline{ $\rightarrow SU(2)_C~, \rightarrow  SU(4),~
  \lambda_{\perp} \to \infty,~ \rightarrow SU(5)$}
 
 \noindent The last scalar is asymptotically free and so the $\lambda_{\perp} \to \infty$ limit can be taken between the SU(4) and SU(5) limits.
 
 An elaborate analysis, involving all reggeon interaction kernels, shows how the complexity of the resulting anomalous wee gauge bosons increases with each infra-red limit. After the SU(2)$_C$ limit, the wee gluons of color superconducting QCD appear. 
With $\lambda_{\perp} \neq \infty $ many fermion loops violate Ward identities and
in the next SU(4) limit left-handed gauge boson interactions (except those mixing with sextet pions) are eliminated. Also left-handed gauge bosons do not contribute as wee gauge bosons. Massive vector bosons, with the flavor symmetry of the sextet pions remain. After $\lambda_{\perp} \to \infty$ and the SU(5) limit, the Critical Pomeron and the massless photon appear together. Also, color octet chiral anomalies at $ k_{\perp} = \infty$ produce bound-states in Standard Model generations.
 
 \subsection{QUD Interactions and States}
 
 The final ``universal wee partons'' are combinations of vector coupling anomalous wee gauge bosons in the adjoint SU(5) representation.
 ``Standard Model'' vector interactions between bound-states appear, that preserve vector SU(3)xU(1)$_{em}$ and couple via anomalies.
 \begin{enumerate}\openup-0.5\jot{
 \item{{\bf The $\tau =  +1$ Critical Pomeron} $\approx$ SU(3) gluon reggeon + wee gauge bosons $\leftrightarrow$ SU(5) singlet projection. There is no BFKL pomeron and no odderon.}
 \item{{\bf The $\tau=-1$ Photon} $\approx$ a U(1)$_{em}$
 gauge boson + wee gauge bosons $\leftrightarrow$ SU(5) singlet projection.}
 \item{{\bf The Electroweak Interaction} $\approx$ left-handed gauge boson mixed with  sextet pion $\leftrightarrow$ SU(5) singlet projection.}}
 \end{enumerate}
 The elementary QUD coupling is kept very small by the infra-red fixed-point and so,  for physical values of the Standard Model couplings to emerge, it is 
crucial that the infinite sums of wee gauge boson anomaly color factors enhance couplings, with 
%\centerline{ 
$$~\alpha_{\scriptscriptstyle QCD} ~> ~\alpha_{\scriptscriptstyle em}~>>~ \alpha_{\scriptscriptstyle QUD} \sim
 \frac{1}{120}$$
 For bound-states, anomaly vertex mixing, combined with fermion and wee parton color
 factors, also produces a wide range of mass scales. 
 
Three Standard Model generations of physical hadrons and leptons appear  
 via  octet anomalies that remain at infinite light-cone momentum after the full SU(5) symmetry is restored. When described in terms of fermions only, all bound-states are SU(5) singlets composed of five elementary fermions, two of which are color octets forming 
 an ``octet pion'' large $k_{\perp}$ anomaly pole contribution. In addition to the color octets, lepton bound states contain three elementary leptons with two producing an anomaly pole. The electron is very close to elementary because the anomaly pole disturbs the Dirac sea minimally. The muon has the same constituents, but in a different anomaly pole dynamical configuration
 that will obviously generate a significant mass. Very significantly, the very small QUD coupling should be the origin of desirably small neutrino masses. 
 Anomaly color factors imply 
 $$M_{\scriptstyle hadrons} >> M_{\scriptstyle leptons} >>  M_{\scriptstyle \nu 's} ~\sim~
  \alpha_{\scriptstyle QUD}$$
  
 Two QUD triplet quark generations give Standard Model hadrons - that mix appropriately. The physical b quark is a mixture\cite{arw6} of all three QUD generations.
 Sextet pions produce electroweak symmetry breaking by mixing with
   left-handed vector bosons. Sextet neutrons \{``neusons''\} are stable (the sextet proton \{``proson''\} is unstable) and so will provide dark matter - with many desirable properties\cite{arw6}.
 Top quark physics is very different from the Standard Model. However, because the sextet ${\eta}$  reproduces the Standard Model final states (at an electroweak scale
 mass~!!) it is hard, experimentally, to distinguish the difference\cite{arw6}. Mixing of triplet and sextet states gives two mixed-parity scalars\cite{arw4,arw5} - the $\eta_3$ and the $\eta_6$. The $\eta_3$ could  be the ``Higgs boson'' discovered at the LHC. The $\eta_6$ may have been seen at the LHC - at the experimental 
  $t\bar{t}$ threshold (where $t$ is the Standard Model top quark). ``Tree-unitarity'' suggests the combined     
  $\eta_3$ and $\eta_6$ couplings should reproduce\cite{arw5} the Standard Model 
  electroweak couplings of the Higgs boson.
 
 There is a significant number of experimental phenomena\cite{arw2,arw3,arw6}, already existing, that provide suggestive (if not definitive) evidence for the QCD sextet quark sector. 
 To see this physics definitively at the LHC, the luminosity should be turned right down\cite{arw3} and the maximum possible energy attained. The detectors should also be modified to cover the maximal possible rapidity range at moderate transverse momentum. It would then be possible to detect the (relatively) large cross-section production of multiple vector bosons, across a wide rapidity range, that is currently missed\cite{arw3,arw6} by the current focus on very large $p_{\perp}$, central rapidity, small cross-section physics. If the Higgs boson is indeed the $\eta_3$ then it will also be produced along with multiple
  vector bosons. It could even be that the production of dark matter neuson/antineuson
  pairs would be seen. Unfortunately, it is likely to take a long time for the needed turnaround in outlook to come to pass.

  \section{The Unique Unitary S-Matrix?}
  
  The infra-red problem of constructing 
  physical states that produce a unitary S-Matrix may actually 
  be more difficult than, and the solution more special 
  and at least as fundamental as, the
  solution of the ultra-violet problem of a field theory via asymptotic freedom.
  I have been led to a very beautiful proposition.
  The relevant entity for particle physics
  is the bound-state S-Matrix of a very special, small $\beta$-function, 
  massless field theory that may only be evident in (and, therefore, need only exist as a quantum field theory in) short-distance perturbation theory. Mass generation becomes an S-Matrix property which is, effectively, separated from the problem of having a sufficiently well-defined short-distance field theory. This could  have the great advantage that (as a matter of principle) there would be no
    need to confront the overwhelmingly difficult, and so far elusively intractable, 
    problem\cite{jw} 
    of constructing a full, non-perturbative, quantum field theory (with or without
    a mass gap) in four dimensions. Note that besides my construction 
    via multi-regge theory, there is no other formalism capable of 
    constructing bound-state scattering amplitudes. Without this ability
    it would not have been possible to envision the existence of an S-Matrix within
    a field theory with the properties of QUD. 
  
   There are  many theoretical virtues for QUD as the origin of the Standard Model,
    including the following. 
   \begin{itemize}\openup-0.5\jot{
   \item{Parity properties of the strong,
   electromagnetic, and weak interactions are naturally explained.}
   \item{ Confinement, chiral symmetry breaking, the parton model and the Critical Pomeron all appear in QCD in a form more consistent with experiment 
   than conventional expectations.}
   \item{ The massless photon partners the ``massless'' Critical $~\pom~$.}
   \item{ Anomaly vertex mixing and wee parton color
   factors produce a wide range of 
   scales and masses, with small neutrino masses due
   to the very small QUD coupling.}
    \item{The only new physics is a high mass sector of the strong
      interaction that gives electroweak symmetry breaking and dark matter.}
   \item{Particles and fields are truly distinct. Physical
   hadrons and leptons have equal status.}
   \item{Symmetries and masses are dynamical S-Matrix
   properties. There are no off-shell amplitudes and there is no Higgs field.}
   \item{As a massless, asymptotically free,  
   fixed-point theory, with no renormalon-related vacuum condensates, QUD induces
   Einstein gravity with zero cosmological constant. 
   However, a QUD S-Matrix without off-shell amplitudes is incompatible with
   quantum gravity. Gravity can not be quantized~!!}}
   \end{itemize}
    
   It would, surely, be incredible 
   if the Standard Model, with all of it's 
   complexity, has the underlying simplicity that I am suggesting. Nevertheless,
   all the necessary ingredients are present and if the predicted effects of the sextet sector are eventually seen clearly at the LHC, I doubt that 
   the radical nature (with respect to the current theory paradigm) 
   of what I am proposing will impede the rapid rise of 
   interest in QUD that will surely ensue. 
   It is important to emphasize again that, in principle,
   there is no freedom for variation in QUD. It is an ``all or nothing'' explanation
   of the origin of the Standard Model. Moreover, it predicts that 
   \begin{center}
   {\large \bf the ``nightmare scenario''- a ``Higgs boson'' produced without new short-distance physics - will occur at the LHC !}
   \end{center}
   
  We can ask, of course, why the underlying massless field theory has to be QUD. My answer would obviously be that I demand the appearance of the Critical Pomeron. However, I can alsoe phrase this requirement in more general terms. The infra-red fixed-point small $\beta$-function is required, firstly 
for the persistence of the scaling wee gluon interactions that enhance infra-red fermion anomaly 
interactions, and
secondly to allow the color-superconductivity starting point that resolves the
quantization ambiguities associated with Gribov copies and Gauss's law. 
The vector interaction non-abelian gauge group has to be as large as 
SU(3) to produce, via scaling wee gluon interactions, a
universal wee gluon distribution that can carry vacuum properties. This property
is surely essential for the existence of an infinite momentum 
``parton model'' that allows asymptotically free perturbation theory to produce an
ultra-violet finite S-Matrix. 

If the vector gauge group is larger than SU(3), 
the anomalous wee gluon scaling interactions are more complicated and
the universal wee parton property is lost. The SU(3) gauge group can, however, be   
extended by left-handed interactions, that aquire a mass via the infra-red anomalies, 
since the only effect
is to also generate bound state masses. This is a beneficial effect in that it
alleviates potential S-Matrix infra-red problems. Asking that this
extension generates masses for all bound-states while introducing no short-distance
anomaly then brings us close to, if not directly to, QUD. Therefore, 
I believe that although I have funneled my discussion 
through the Critical Pomeron, in fact all of the properties needed to obtain a
well-defined particle S-Matrix may come together to uniquely select QUD. If so,
\begin{center}
{\large \bf the Standard Model could be reproducing the Unique, Unitary, S-Matrix !!!}
\end{center}

\end{document}